\documentclass[prl,letterpaper,twocolumn,showpacs,superscriptaddress,floatfix]{revtex4}
\usepackage{graphicx,psfrag,amsmath,amssymb,amsfonts,bbm,latexsym,color,dcolumn,bm}

\begin{document}
\title{Optimal cooling strategies for magnetically trapped atomic Fermi-Bose mixtures}

\author{Michael Brown-Hayes}
\affiliation{Department of Physics and Astronomy, Dartmouth College, 6127 Wilder Laboratory, 
Hanover, NH 03755}

\author{Roberto Onofrio}
\affiliation{Department of Physics and Astronomy, Dartmouth College, 6127 Wilder Laboratory, 
Hanover, NH 03755}
\affiliation{Dipartimento di Fisica ``G. Galilei'', Universit\`a di Padova, Via Marzolo 8, Padova 35131, Italy}
\affiliation{Center for Statistical Mechanics and Complexity, INFM, Unit\`a di Roma 1, Roma 00185, Italy}


\begin{abstract}
We discuss cooling efficiency for different-species Fermi-Bose mixtures in magnetic traps. 
A better heat capacity matching between the two atomic species is achieved by a proper 
choice of the Bose cooler and the magnetically trappable hyperfine
states of the mixture. 
When a partial spatial overlap between the two species is also taken into
account, the deepest Fermi degeneracy is obtained for an optimal 
value of the trapping frequency ratio. 
This can be achieved by assisting the magnetic trap with a 
deconfining light beam, as shown in the case of fermionic 
${}^6$Li mixed with ${}^{23}$Na, ${}^{87}$Rb, and ${}^{133}$Cs, 
with optimal conditions found for the not yet explored 
${}^6$Li-${}^{87}$Rb mixture.
\end{abstract}

\pacs{05.30.Fk, 32.80.Pj, 67.60.-g, 67.57.-z}
\maketitle
The availability of new cooling techniques in the nanokelvin range has
recently opened up a new field of study of condensed matter systems at
low densities.  
Mean-field theory can be reliably used in the low-density
limit to understand many-body quantum phenomena involving dilute
Bose gases, in particular superfluidity \cite{Reviewsbooks}.  
In the case of trapped Fermi gases, which maps the physics of
interacting electrons in a metal, impressive progress has been 
made in the last few years.  
Following the first observation of quantum degeneracy for an 
atomic Fermi gas \cite{DeMarco}, both ideal behaviour - the effect 
of the Fermi pressure \cite{Truscott,Schreck} - and interacting 
features in Fermi-Bose mixtures \cite{Inguscio} and in two-component 
Fermi gases \cite{Ohara}, have been reported. 
More recently, Bose-Einstein condensation of tightly confined 
Fermi pairs have been observed by various groups
\cite{Regal,Bartenstein,Zwierlein,Kinast,Bourdel}, including the 
spectroscopic results \cite{Chin} which have been interpreted as the 
first evidence for a pairing gap in a fermionic superfluid \cite{Kikkunen}. 
These latest results are based upon the manipulation of the fermions 
scattering length through Feshbach resonances, in a strong
coupling regime for the interacting atoms, also named {\sl resonant} 
superfluidity \cite{Resonant}. 
While this evidence is a major breakthrough towards 
understanding the BCS-BEC cross-over, it is also important to observe 
such a transition in a system with moderate scattering length, as 
this will mimic in a different context the situation of BCS 
Cooper pairs with the large correlation distance typical of 
low-$\mathrm{T_c}$ superconductors \cite{Stoof}. 
According to theoretical predictions, this requires reaching much 
lower temperatures than the ones at which resonant superfluidity is
expected, implying more challenges from the experimental viewpoint. 
In this paper we discuss the efficiency for sympathetically cooling  
different-mass Fermi-Bose species in a magnetic trap. 
We identify an optimal range of parameters 
where  heat capacity matching between the two species in the
mixture is obtained while optimizing their spatial overlap. 
It turns out that, by using a light-assisted magnetic trap -  
a feasible addition to many experimental setups already 
available for Fermi cooling - degeneracy parameters 
$T/T_\mathrm{F}$ in the $10^{-3}$ range could be achieved.

In order to assess more quantitatively the cooling strategies, we consider 
magnetically trapped Fermi-Bose mixtures made of fermionic ${}^6$Li, 
a promising candidate for observing superfluidity features
\cite{Stoof}, and bosonic ${}^{23}$Na, ${}^{87}$Rb, and ${}^{133}$Cs. 
Our discussion is elicited by a recent result obtained at MIT 
\cite{Hadzibabic1}, where a value of $T/T_\mathrm{F} \simeq 0.05$ 
has been achieved by using a ${}^6$Li-${}^{23}$Na mixture in a magnetic trap. 
Other groups using same species Fermi-Bose mixtures like ${}^6$Li-${}^7$Li
have instead obtained higher $T/T_{\mathrm F}$ \cite{Truscott,Schreck}. 
The MIT result can be simply explained, as discussed 
in \cite{jsp}, by considering the heat capacities of ideal Bose and
Fermi that can be approximated in the degenerate regime respectively as:
\begin{equation}
C_\mathrm{B} \simeq 10.8 ~ k_B N_B
\left(\frac{T}{T_\mathrm{c}}\right)^3, \ \ \
C_\mathrm{F} \simeq  \pi^2 ~ k_B N_F \frac{T}{T_\mathrm{F}}, 
\label{heatcapacities}
\end{equation}
where $T$ is the temperature, $T_\mathrm{c}$ and $T_\mathrm{F}$ the 
critical temperature for the Bose-Einstein phase transition and the 
Fermi temperature respectively, $N_B$ and $N_F$ the corresponding 
number of trapped atoms, and $K_B$ the Boltzmann constant.
By considering the heat capacity ratio $C_\mathrm{B}/C_\mathrm{F}$, 
supposing that there is complete spatial overlap between the 
two species,  and expressing $T_c$ and $T_\mathrm{F}$ in terms of the trapping 
frequencies $\omega_\mathrm{B}$ and $\omega_\mathrm{F}$, 
the $T/T_\mathrm{F}$ ratio can be written as: 
\begin{equation}
\frac{T}{T_\mathrm{F}} \simeq 0.35 
\left(\frac{\omega_B}{\omega_F}\right)^{3/2} 
\left(\frac{C_B}{C_F}\right)^{1/2}.
\label{degeneracyparam}
\end{equation}
Since the MIT experiment involves a relatively large mass ratio
between the two species, $m_B/m_F \simeq 4$, and the trapping 
strengths are similar in a magnetic trap, the trapping frequency 
ratio $\omega_B/\omega_F \simeq (m_F/m_B)^{1/2}$ implies a lower 
$T/T_F$ degeneracy parameter with respect to the case of ${}^6$Li 
sympathetically cooled with ${}^7$Li. 
By assuming that sympathetic cooling stops completely when $C_B/C_F
\simeq 0.1$, an hypothesis which should be close to reality for
experiments with complete evaporation of the Bose component as 
in \cite{Hadzibabic1}, we obtain $T/T_F \simeq 0.11$ for a 
${}^6$Li-${}^7$Li mixture, while $T/T_F \simeq 0.04$ for 
a ${}^6$Li-${}^{23}$Na mixture, estimates not dissimilar from 
the actual experimental values. 

The fact that a large mass ratio between the Bose and the Fermi species is 
beneficial for reaching lower temperatures is due to basic quantum statistical laws. 
As seen from Eq. (\ref{heatcapacities}), the specific heat of a Bose 
gas in the degenerate regime decreases with the temperature as $T^3$, 
while a Fermi gas has a milder, linear, temperature dependence. 
In order to preserve a large heat capacity it is therefore important to 
maintain the Bose gas as close as possible to the classical regime. 
More classicality for the Bose gas can be easily achieved by using 
more massive species or, equivalently, in an inhomogeneous situation 
such as the one of harmonic trapping, by having smaller vibrational 
quanta $\hbar \omega_B$ \cite{note1}. 
A purely optical trapping solution has been proposed in
\cite{Onofrio}, by selectively trapping the two species in different 
potentials via different detunings in a bichromatic optical trap. 
In the context of magnetic trapping this can be instead obtained by
using larger mass ratios $m_B/m_F$, for instance using ${}^{87}$Rb 
or ${}^{133}$Cs as Bose coolers, and by properly choosing the combination 
of trapped hyperfine states. 
Different hyperfine states have different magnetic moments, and in
Table I we show possible hyperfine states for various combinations of  
mixtures which are trappable with weak magnetic fields, including
their trapping frequency ratio. 
\begin{table}
\begin{ruledtabular}
\begin{tabular}{cccc}
 Fermi-Bose mixture    & hyperfine states      & $\alpha$  & $\omega_F/\omega_B$ \\ 
\hline  
 ${}^6$Li-${}^{23}$Na  & Li(1/2,-1/2)-Na(1,-1) & $2/3$        &       $1.599$       \\
                       & Li(1/2,-1/2)-Na(2,2)  & $1/3$        &       $1.130$       \\
                       & Li(3/2,3/2)-Na(1,-1)  & $2$          &       $2.769$       \\
                       & Li(3/2,3/2)-Na(2,2)   & $1$          &       $1.958$       \\
\hline 
${}^6$Li-${}^{87}$Rb   & Li(1/2,-1/2)-Rb(1,-1) & $2/3$        &       $3.109$       \\ 
                       & Li(1/2,-1/2)-Rb(2,2)  & $1/3$        &       $2.198$       \\
                       & Li(3/2,3/2)-Rb(1,-1)  & $2$          &       $5.385$       \\ 
                       & Li(3/2,3/2)-Rb(2,2)   & $1$          &       $3.808$       \\
\hline
${}^6$Li-${}^{133}$Cs  & Li(1/2,-1/2)-Cs(3,-3) & $4/9$        &       $3.138$       \\
                       & Li(1/2,-1/2)-Cs(4,4)  & $1/3$        &       $2.718$       \\    
                       & Li(3/2,3/2)-Cs(3,-3)  & $4/3$        &       $5.436$       \\      
                       & Li(3/2,3/2)-Cs(4,4)   & $1$          &       $4.707$       \\ 
\end{tabular}
\end{ruledtabular}
\caption{Magnetically trappable Fermi-Bose mixtures in the ground state 
with ${}^6$Li as the fermionic component. For each mixture 
we report the hyperfine state $(F,m_{\mathrm F})$, the ratio between 
the $m_F g_F$ factors of the Fermi and the Bose hyperfine states, 
$\alpha=(m_F g_F)_\mathrm{fermion}/(m_F g_F)_\mathrm{boson}$,  
and the corresponding trapping frequency ratio.} 
\end{table}
It is evident that trapping frequency ratios $\omega_F/\omega_B$ up to
almost 5.5 can be obtained by using particular hyperfine states in
the case of Rubidium and Cesium. However, there are two issues 
to be taken into account in the analysis when realistic losses 
of particles unavoidably occurring in the trap are taken into account. 
Firstly, some mixtures are not optimal in terms of spin-exchange
losses, and this favors some stretched states which have a natural 
protection against such a kind of losses.   
Secondly, very large mass ratios imply diminished spatial overlap 
between the trapped clouds, and also larger relative sagging due 
to the gravitational field. The consequent lack of complete overlap 
between the two species will affect the cooling efficiency. 
In the extreme case of very large mass ratio for the two species 
the two atomic clouds, due to gravity sagging, could end up with no overlap 
at all therefore completely inhibiting any cooling. 
In the presence of losses inducing heating, as theoretically 
discussed in \cite{Timmermans,Castin,Dziarmaga}, the final 
temperature of the mixture will depend upon the balance between 
cooling and heating rates, and the partial overlap will diminish 
the former, shifting the equilibrium temperature to higher values 
than those estimated in Eq. (2). 

The degree of spatial overlap between the Fermi and the Bose clouds 
in the inhomogeneous case can be quantified by considering a 
dimensionless figure of merit defined as \cite{note2}:
\begin{equation}
\eta=\int \rho_F^{1/2}(r) \rho_B^{1/2}(r) d^3 r,
\end{equation}
where $\rho_F$ and $\rho_B$ are the densities of the Fermi and the
Bose gases. 
From a quantum mechanical viewpoint $\eta$ represents the scalar
product between the amplitudes of the macroscopic wavefunctions
associated with the two gases. 
The quantity $\eta^2$ is proportional to the number of atoms 
which share the same region of space and therefore the cooling 
rate in the presence of a partial overlap will be decreased by the 
geometric factor $\eta$ with respect to ideal overlap as 
$\dot{Q}_\mathrm{cool} \rightarrow \eta^2 \dot{Q}_\mathrm{cool}$. 
Accordingly, the minimum attainable degeneracy parameter 
$T/T_{\mathrm F}$ in the presence of a partial overlap will 
be increased as $T/T_{\mathrm F} \rightarrow \eta^{-2} T/T_{\mathrm
  F}$, since the balance between heating rate and cooling rate 
is now shifted towards higher equilibrium temperatures. 
The key point is that by choosing different trapping frequencies 
for the Bose and the Fermi gases one can match their spatial sizes 
in spite of a large mass ratio between the two species. 
In particular, since the Bose gas is more massive and therefore more 
confined in a purely magnetic trap, its confinement should be made
less stiff in order to match the size of the more delocalized, lighter
Fermi gas. 

A simple analysis can be developed in the Thomas-Fermi approximation, 
as in this case the density profiles of the two species can be
obtained from the knowledge of the local chemical potentials. 
In a zero-temperature homogeneous Fermi-Bose mixture the chemical
potentials $\mu_F, \mu_B$ are written in terms of the densities
$\rho_F, \rho_B$ as \cite{Molmer,Roth}:
\begin{eqnarray}
\mu_B & = & \lambda_B \rho_B+ \lambda_{FB} \rho_F
\nonumber \\ 
\mu_F & = & (6\pi^2)^{2/3} \frac{\hbar^2}{2 m_F} \rho_F^{2/3}+ \lambda_{FB}
\rho_B,
\label{potential}
\end{eqnarray}
where the parameters $\lambda_B, \lambda_{FB}$ have been introduced, with
$\lambda_i= 8 \pi \hbar^2 a_i/m$, $a_i$ ($i=B$ or $FB$) being the
s-wave elastic scattering length.
In the inhomogeneous case corresponding to trapping with potentials
$V_B(r), V_F(r)$ we have $\mu_B \rightarrow \mu_B-V_B(r)$,
$\mu_F\rightarrow \mu_F-V_F(r)$, and then:
\begin{eqnarray}
\rho_B(r) & = &  \frac{\mu_B-V_B(r)}{\lambda_B}- 
\frac{\lambda_{FB}}{\lambda_B} \rho_F(r)
\nonumber \\ 
\rho_F(r) & = & \frac{1}{6 \pi^2 \hbar^3} \{2 m_F [\mu_F-V_F(r)] -
2 m_F \lambda_{FB} \rho_B \}^{3/2}, \ \ 
\label{densities}
\end{eqnarray}
where the density dependence in the case of null interspecies interaction is 
easily recovered for $\lambda_{FB}=0$. 
An iterative technique in which the trial initial densities are the
ones corresponding to non-interacting species allows us to evaluate 
numerically the density profiles in a three-dimensional situation 
also taking into account the presence of gravity. 
In Fig. 1 we report the degeneracy parameter $T/T_F$ versus the 
$\omega_F/\omega_B$ ratio for the stretched states of the mixtures, 
corresponding to the bottom row for each mixture in Table I. 
Initially the degeneracy parameter has an inverse dependence upon 
$\omega_F/\omega_B$,  as expected on the basis of Eq. (\ref{degeneracyparam}) 
and the fact that the size of the heavier Bose gas is approaching 
that of the lighter Fermi counterpart. 
When the trapping frequency ratio is such that the Bose gas size 
exceeds the size of the Fermi gas, and the relative sagging is starting 
to play a role due to the weakened trapping frequency for the Bose
gas, the overlap factor decreases significantly and the $T/T_F$ ratio
increases. An optimal $\omega_F/\omega_B$ is then obtained for each
mixture, with the lowest values obtained for the $^6$Li-$^{87}$Rb 
mixture as a compromise betweeen a larger ratio $m_B/m_F$ 
with respect to the $^6$Li-$^{23}$Na mixture, and a smaller 
relative sagging with respect to the $^6$Li-$^{133}$Cs mixture.
In the latter mixture, unlike the former two mixtures, the effect of 
sagging is quite substantial, as it can be seen in Fig. 2 where the 
degeneracy parameter is shown versus the $\omega_F/\omega_B$ ratio 
in presence or absence of gravity, for various values of the 
absolute trapping frequency of the Bose species. 
Indeed, gravity sagging introduces an absolute scale for the
frequencies in such a way that the cooling curves depend 
not only on the $\omega_\mathrm{F}/\omega_\mathrm{B}$ ratio, 
but on the absolute values of the frequencies. 
Analogous curves for the other two mixtures show a much 
milder dependence on gravity sagging.
\begin{figure}[t]
\psfrag{x}[][]{$\omega_\mathrm{F}/\omega_\mathrm{B}$}
\psfrag{y}[][]{$T/T_\mathrm{F}$}
\includegraphics[width=1.00\columnwidth,clip]{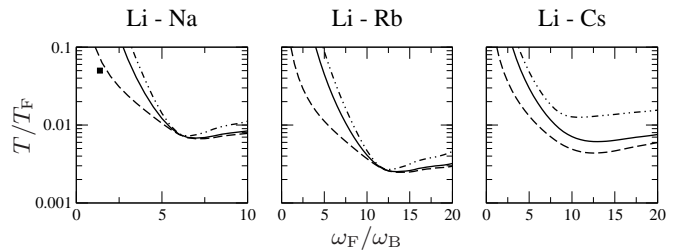}
\caption{Optimal sympathetic cooling of ${}^6$Li with different Bose
  coolers. The degeneracy parameter $T/T_\mathrm{F}$ is shown 
versus the trapping frequency ratio $\omega_F/\omega_B$ for 
the cases of fermionic ${}^6$Li mixed with bosonic ${}^{23}$Na 
(left), ${}^{87}$Rb (center), and ${}^{133}$Cs (right). 
The various curves within each mixture refer to various 
possible values for the interspecies scattering length, 
$a_{FB}=-0.5$nm (dashed line), $a_{FB}=0.0$nm (continuous), 
$a_{FB}=1.0$nm (dot-dashed), while the known intraspecies scattering lengths 
for the Bose components have been assumed.  
The square corresponds to the MIT result obtained for the 
$(3/3,3/2)-(2,2)$ combination of the ${}^6$Li-${}^{23}$Na 
mixture \cite{Hadzibabic1}.}
\label{fig1}
\end{figure}
The various curves for each Fermi-Bose mixture in Fig. 1 correspond to different values for 
the interspecies scattering length, including the case without interspecies interactions. 
It is worth pointing out that the interspecies scattering lengths have
not yet been measured, and that our analysis reproduces the observed value
for the degeneracy parameter obtained by the MIT group if a negative 
value $a_{FB} \simeq -(0.5 \div 1)$nm is assumed, at variance with the 
theoretical estimate reported in \cite{Timmercote}, and compatible 
with the discussion reported in \cite{Roth} where a window on the 
values of the interspecies scattering length was estabilished 
based on issues of stability and lack of phase separation for the
Fermi-Bose mixture. 
The presence of attractive interspecies interactions ($a_{FB}<0$), 
naturally present or momentarily induced by a Feshbach resonance, 
could allow to increase for some time the overlap between the species 
enhancing the cooling speed, as demonstrated in the case of the 
${}^{40}$K-${}^{87}$Rb mixture \cite{Ferrari}.
This analysis is limited by the static picture implicit in
Eq. (\ref{degeneracyparam}), which in a more detailed study should 
be superseded by a simulation of the master equation for the
Fermi-Bose mixture during cooling, also including the dependence 
of the density profiles upon the temperature \cite{Amoruso}.   
Despite these limitations, from Fig. 1 it is  evident that the
magnetic trap alone is far from being optimized for cooling, and 
that more selective confinement of the Fermi and the Bose species 
is necessary to achieve the highest degeneracy. 

\begin{figure}[t]
\psfrag{x}[][]{$\omega_\mathrm{F}/\omega_\mathrm{B}$}
\psfrag{y}[][]{$T/T_\mathrm{F}$}
\psfrag{a}[][]{a}
\psfrag{b}[][]{b}
\psfrag{c}[][]{c}
\includegraphics[width=0.90\columnwidth]{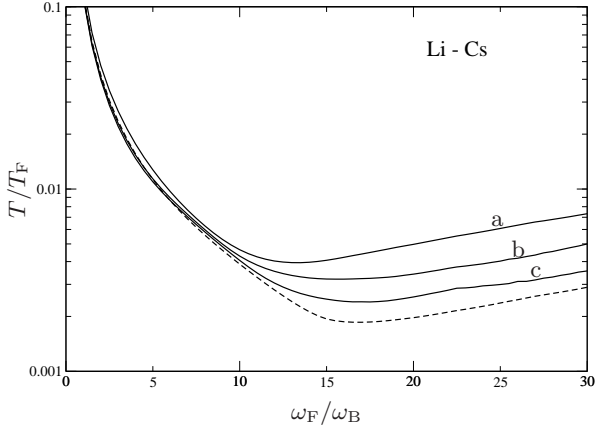}
\caption{Optimal sympathetic cooling for a ${}^6$Li-${}^{133}$Cs 
in presence of gravity. The degeneracy parameter $T/T_\mathrm{F}$ is shown 
versus the trapping frequency ratio $\omega_F/\omega_B$ for various 
values of the absolute frequency of the Bose species, in absence
(dashed) and in presence of gravity 
(continuous curves, with  
$\omega_\mathrm{B}/2\pi=50$Hz (a), 
$\omega_\mathrm{B}/2\pi=100$Hz (b), 
$\omega_\mathrm{B}/2\pi=150$Hz (c)), 
for an interspecies scattering length of $a_{FB}=-0.5$nm. 
The influence of gravity sagging in the cooling efficiency 
is particularly evident for weak confinement.}
\label{fig2}
\end{figure}

One way to move the trapping frequency ratio in the region
where $T/T_F$ is minimized is to use a focused laser beam 
as in usual optical dipole traps. The different detunings 
of the light with respect to the atomic transitions can be 
exploited to achieve a weaker confinement for the Bose species. 
Superimposed with the trapping magnetic potential, generated for
instance through a Ioffe-Pritchard magnetic trap, a laser beam
propagates along the axial direction of the magnetic trap. 
The trapping angular frequencies in the presence of the
laser beam can be written as ($\alpha=\mathrm{B},\mathrm{F}$):
\begin{equation}
\omega_\mathrm{\alpha}=
{[(\omega_{\alpha}^{ax\,2}+\xi_{\alpha}^{ax\,2})^{1/2}
 (\omega_{\alpha}^{r\,2}+\xi_{\alpha}^{r\,2})]}^{1/3},
\label{assistedtrap}
\end{equation}
where $\omega_{\alpha}^{ax}$ ($\omega_{\alpha}^{r}$) is the intrinsic 
trapping angular frequency of the magnetic trap for the species
$\alpha$ in the axial (radial) direction and $\xi_\alpha^{ax}$  
($\xi_\alpha^r$) is the angular frequency due to the focused 
laser beam in the axial (radial) direction. The latter are derived 
by the trapping potential due to the laser beam, which we assume to 
propagate along the direction $x$: 
\begin{figure}[b]
\psfrag{x}[]{P (mW)}
\psfrag{y1}[]{$\omega_\mathrm{F}/2\pi,\omega_\mathrm{B}/2\pi$ (Hz)}
\psfrag{y2}[]{$\omega_\mathrm{F}/\omega_\mathrm{B}$}
\includegraphics[width=1.0\columnwidth,clip]{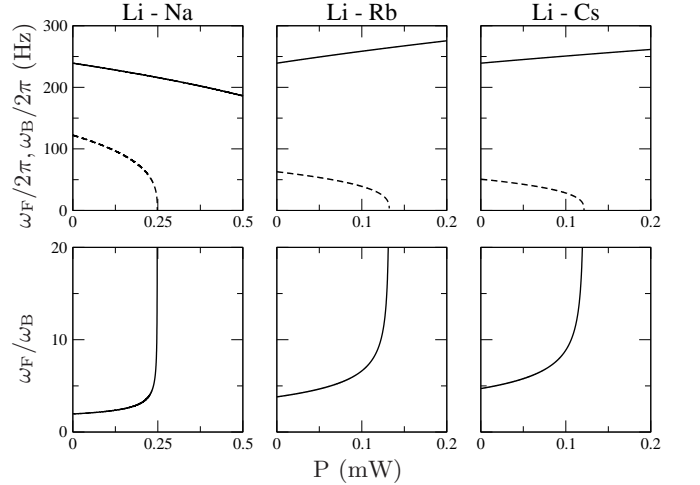}
\caption{Trapping Fermi-Bose mixtures in light-assisted magnetic traps.
The trapping frequencies (top panels, Bose species dashed line) 
and their ratio (bottom panels) are depicted versus the laser power 
of the deconfining beam for the Bose species, for the cases of 
${}^{6}$Li-${}^{23}$Na (left), ${}^{6}$Li-${}^{87}$Rb (center), 
and ${}^{6}$Li-${}^{133}$Cs (right). 
We discuss the case of a Ioffe-Pritchard magnetic trap with radial 
gradient $B_r'=170$ G/cm, axial curvature $B_{ax}^{''}=125$ G/cm${}^2$, 
and bias field $B_0$=1 G. The optical antitrap is assumed to have 
a focused laser beam with waist $w=8\mu$m, and a wavelength 
$\Lambda$= 560, 741, and 808 nm in the three cases, respectively. 
These wavelengths are blue-detuned by 5$\%$ with respect to the 
corresponding atomic transition for the Bose species, enough to 
ensure small residual scattering rates.}
\label{fig3}
\end{figure}
\begin{equation}
U_\alpha(x,r)=
-\frac{\hbar \Gamma_\alpha^2 T^\alpha P}{4 \pi I_\alpha^{\mathrm sat}w^2} 
\frac{\exp
  \left(\frac{-2r^2}{w^2\left(1+x^2/R^2\right)}\right)}{1+x^2/R^2},
\label{potential}
\end{equation}
where $r^2=y^2+z^2$ in the radial direction, 
$T^\alpha=1/(\Omega_\alpha-\Omega)+1/(\Omega_\alpha+\Omega)$ 
is a parameter related to the detuning between the atomic transition angular 
frequencies $\Omega_\alpha=2 \pi c/\Lambda_\alpha$, and the laser angular 
frequency $\Omega=2 \pi c/\Lambda$ ($\Lambda_\alpha$, $\Lambda$ being 
the atomic transition and laser beam wavelengths, respectively), $P$ and $w$  
power and waist of the laser beam, $R=\pi w^2/\Lambda$ its Rayleigh
range, $\Gamma_{\alpha}$ the atomic transition linewidth, 
and $I_\alpha^{\mathrm sat}=\hbar \Omega_\alpha^3 
\Gamma_\alpha/12 \pi c^2$ is the saturation intensity for 
the atomic transition.
By setting $\hbar \Gamma_\alpha^2 T^\alpha/{4 \pi I_\alpha^{\mathrm
    sat}w^2}=\delta_\alpha$ 
we get contributions to the angular trapping frequencies due 
to the deconfining beam as:
\begin{eqnarray}
\xi_{\alpha}^{ax} & = &  
\sqrt{\frac{1}{m_\alpha} \frac{\partial^2 U}{\partial x^2}}=                   
\sqrt{\frac{2 \delta_\alpha P}{m_\alpha R^2}}
\nonumber \\
\xi_{\alpha}^r & = & 
\sqrt{\frac{1}{m_\alpha} \frac{\partial^2 U}{\partial y^2}}=                   
\sqrt{\frac{1}{m_\alpha} \frac{\partial^2 U}{\partial z^2}}=                   
\sqrt{\frac{4 \delta_\alpha P}{m_\alpha w^2}}.
\label{frequencies}
\end{eqnarray}
If the laser wavelength $\Lambda$ is blue-detuned with respect to 
the atomic transition wavelength of the Bose species the contribution 
of the focused laser beam will decrease the curvature of the overall 
trapping, corresponding to imaginary values for  
$\xi_{\alpha}^{ax}, \xi_{\alpha}^r$ in Eq. (\ref{frequencies}).  
In Fig. 3 we present the dependence of the frequencies and their 
ratio versus the laser power for a fixed relative detuning of the
laser beam with respect to the atomic transition wavelength of the
Bose species. 
Due to the smaller atomic transition wavelength for ${}^6$Li versus 
${}^{23}$Rb and ${}^{133}$Cs, the trapping strength of the 
Fermi species is also increased, while for the case of ${}^{23}$Na 
there is a slight decrease. It is evident that for moderate values of the 
power of the laser beam it is possible to bring the $\omega_F/\omega_B$ 
ratio to the value minimizing the degeneracy parameter
$T/T_\mathrm{F}$ for the three mixtures. 
Although we have focused on a particular example, the possibility 
to deconfine in a selective way a magnetically trapped species 
by means of a focused blue-detuned beam is quite general, and 
relies on well explored techniques which have been implemented 
for other purposes, namely the suppression of Majorana spin-flip losses
during evaporative cooling of Bose species \cite{Davis} (see also 
\cite{Raman} for related recent work). 
 
In summary, we have discussed Fermi-Bose cooling in a magnetic 
trap and how it can be optimized by using a laser beam.
The result of this analysis can be summarized as showing that 
the ${}^6$Li-${}^{87}$Rb mixture is the more promising for 
achieving degeneracy parameters in the $10^{-3}$ range, provided 
that the interspecies scattering length is large enough and 
possibly negative. This also leads to the need for determining 
the interspecies scattering lengths, as this 
is crucial to identify configurations manifesting efficient
sympathetic cooling, possibly also exploiting Feshbach resonances 
at low or moderate values of magnetic field.
With respect to the ${}^6$Li-${}^{23}$Na mixture, ${}^6$Li-${}^{87}$Rb
allows for significantly larger $\omega_\mathrm{F}/\omega_\mathrm{B}$
ratios, and consequently smaller values of $T/T_\mathrm{F}$, could be achieved.
Over the ${}^6$Li-${}^{133}$Cs, ${}^{87}$Rb has the advantage that 
the relative gravitational sagging is less pronounced, and it 
is easier to bring close to Bose condensation. 
Moreover, on the practical side, both Lithium and Rubidium can 
take advantage of the same diode laser technology. 
The achievement of $T/T_\mathrm{F} \simeq 10^{-3}$ is considered 
crucial to obtain fermion superfluidity in the widest range of the 
relevant parameter space and to study Fermi-Bose quantum fluids \cite{Mixtures}.

\begin{acknowledgments}
MBH acknowledges support from the Dartmouth Graduate Fellowship
program, and RO acknowledges partial support through Cofinanziamento 
MIUR protocollo 2002027798$\_$001. 
\end{acknowledgments}


\begin{thebibliography}{99}
\bibitem{note2} The choice of other indicators for the degree of
  spatial overlap, such as the product of the two densities then
  integrated over space,  $\int d^3 r \int d^3r' \rho_F(r) \rho_B(r')$, 
does not change the analysis within a 10$\%$ accuracy.
\end{thebibliography}
\end{document}